# Integer, fractional and side band injection locking of spintronic feedback nano-oscillator to microwave signal

Hanuman Singh[1], K. Konishi[2], S. Bhuktare[1], A. Bose[1], S. Miwa[2], A. Fukushima[3], K. Yakushiji[3], S. Yuasa[3], H. Kubota[3], Y. Suzuki[2], A. A. Tulapurkar[1]

[1]Department of Electrical Engineering, Indian Institute of Technology Bombay, Powai, Mumbai – 400 076, India
[2]Graduate School of Engineering Science, Osaka University, Toyonaka, Osaka -560-8531, Japan
[3]National Institute of Advanced Industrial Science and Technology (AIST), Spintronics Research Center, Ibaraki -305-8568, Japan

*In this article we demonstrate the injection locking of recently demonstrated spintronic feedback nano oscillator to microwave magnetic fields at integers (n=1, 2, 3) as well fractional multiples (f=1/2, 3/2 and 5/2) of its auto oscillation frequency. Feedback oscillators have delay as a new "degree of freedom" which is absent for spin-transfer torque based oscillators, which gives rise to side peaks along with a main peak. We show that it is also possible to lock the oscillator on its side band peaks, which opens a new avenue to phase locked oscillators with large frequency differences. We observe that for low driving fields, sideband locking improves the quality factor of the main peak, whereas for higher driving fields the main peak is suppressed. Further, measurements at two field angles provide some insight into the role of symmetry of oscillation orbit in determining the fractional locking.*

Nano-oscillators based on the phenomenon of spin-transfer torque (STT) [1-7] have been the subject of considerable research due to their potential applications in microwave communication devices as well as due to the rich physics involved in their operation. Recently our group demonstrated a spintronic feedback nano-oscillator (SFNO) [8,9] based on the tunneling magneto-resistance (TMR) effect and works without STT. The working principle of SFNO is as follows: SFNO comprises a magnetic tunnel junction (MTJ) nano-pillar connected to a waveguide on top of it. Such a system, when powered by dc current can amplify rf signals: Rf current passing through the waveguide excites the magnetization of the free layer via oscillating Oersted magnetic field, and the dc current passing through MTJ converts the oscillations of magnetization into ac voltage via tunneling magneto-resistance (TMR) effect. Above a certain dc current level, the input signal is amplified. Thus the free layer of MTJ apart from being a resonator can also work as an amplifier. If we connect a feedback path from MTJ to the waveguide, the system works as an oscillator.

It was demonstrated that SFNO [9] can exhibit very large quality factors exceeding 10,000. However, to obtain large output power, it is necessary to connect many oscillators together. If many oscillators can be synchronized, i.e. they oscillate with the same frequency and phase [10], we can get larger power output. We therefore studied the injection locking of a single SFNO to a microwave source, which acts as the second oscillator. Further such studies are also important to understand the complex non-linear dynamics of an auto-oscillator and its possible applications to neuromorphic computing [11-13]. Injection locking of spin-transfer nano-oscillator (STNO) [14-23] has been the subject of many experimental and theoretical investigations with the same motivation. SFNO is not limited to the feedback from Oersted magnetic field, but can be even realized from spin current generated by using inverse Spin-Hall effect [24-25]. Even interfacial Rashba coupling can be used as a feedback mechanism [24]. A large Rashba coupling has been demonstrated recently [26, 27], which could be potentially used for making oscillators. In this case, the mathematical treatment of the oscillator is similar to the Oersted field feedback and experimental results obtained here would be applicable to such interfacial Rashba coupling driven oscillators. Further, it is possible to improve quality factor and output power of the oscillator by combining the magnetic field feedback and STT effects in the same MTJ [28], which makes the exploration of the feedback effect even more attractive.

Here we demonstrate that SFNO can be locked to external microwave magnetic field at integer as well as fractional multiples of its auto-oscillation frequency. We carried out these studies for two directions of external dc magnetic field, which sheds some light on the relation between locking range and symmetry of oscillation with respect to the external driving magnetic field. As SFNO is a delay line oscillator, its power spectrum shows a main peak accompanied by side peaks. We found that the side peaks also show phase locking phenomenon. Thus oscillators with large difference in the auto-oscillation frequencies can be phase locked using side bands. The side band separation can be controlled by the delay, which gives us a new handle to lock the phase of oscillators.

We fabricated an MTJ stack on thermally grown $SiO_2$ (500 nm) with the following structure: Ta(5)/ Cu(20)/ Ta(5)/ Cu(20) / Ta(3) / Ru(5) / IrMn(7) / CoFe(3) / Ru(0.8) / CoFeB(3) / CoFe(0.4) / MgO(0.9) / CoFeB(3) /Ta(5) / Cu(30)/ Ta(5)/ Ru(5) (numbers in bracket denotes the thickness in nm). The elliptical nano pillars of size 300 x 500 $nm^2$ were fabricated using e-beam lithography and Ar ion milling from the multilayer stack. A co-planar waveguide (CPW) with about 1 μm width was fabricated on top of the MTJ nano-pillar and is electrically insulated from the MTJ by a 100 nm thick $SiO_2$ layer. The CPW is oriented in such a way that the current passing through it creates a magnetic field along the x-direction. The experimental arrangement is shown in Fig. 1(a). The fixed layer magnetization is along x-axis which is also the easy axis of the free layer. The external magnetic field is applied along the y-axis. A constant bias current was passed through the MTJ using a bias-tee network. The RF voltage, generated across the MTJ due to the oscillation of the magnetization direction of the free layer is divided into two parts: one part is measured by a spectrum analyzer, and the other part is fed into the CPW after passing it through an amplifier. As argued in ref.[ 9], an amplifier is not required for a "perfect" device with larger TMR and narrower feedback line. The oscillating current passing through the CPW, creates an oscillating magnetic field. This oscillating magnetic field acts as the feedback signal and amplifies the oscillation of the free layer magnetization. Thus, the MTJ with feedback connection as shown in Fig. 1(a) functions as an oscillator. To study the locking of the oscillator to external rf magnetic field, rf current was injected from a source into the feedback path through a directional coupler as shown in Fig. 1(a). All the experiments were carried out at room temperature.

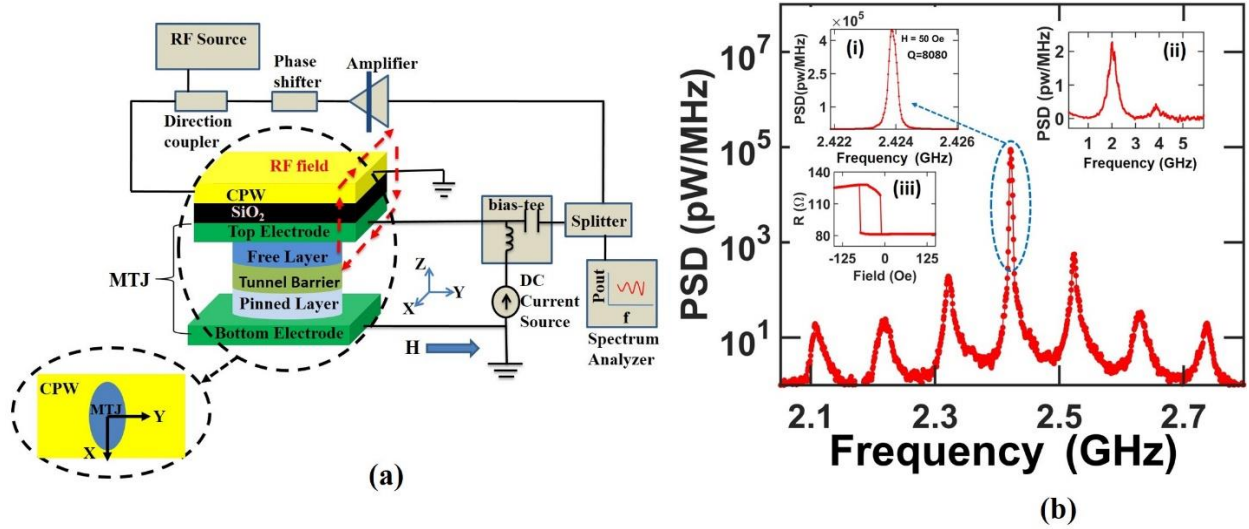

*FIG.1. (a) Schematic diagram for the injection locking of spintronic feedback nano oscillator. The MTJ consists of the free layer, the tunnel barrier and the pinned layer. On the top of MTJ a coplanar wave guide is situated which is electrically insulated from MTJ. MTJ is biased by passing a DC current through bias-tee and we get oscillating voltage due to oscillation in the magnetization of the free layer. Output signal is split into two parts using power splitter. One part is amplified and fed back to CPW. External rf signal is added to the feedback signal by using a directional coupler. The rf current passing through CPW creates a microwave field which changes the dynamics of the free layer. The second part of the output signal is observed on a spectrum analyzer. The inset in Fig.1(a) shows the orientation of MTJ nano pillar on CPW. Fig. 1(b) shows Power spectral density of the free running oscillator (i.e. MTJ with feedback but without external locking magnetic field) in $\log_{10}$ scale, with applied magnetic field of 50 Oe along y-axis, bias current of -2.3mA and 27 dB gain of amplifier. The first inset shows the same zoomed-in spectrum in linear scale. The main peak shows a high Q factor of 8080. The second inset shows the noise spectrum obtained for the same magnetic field of 50 Oe along y-axis and same bias current of -2.3mA by disconnecting feedback line from CPW. The third inset of Fig. 1(b) shows the TMR of the device with magnetic field swept along easy axis (X- axis).*

The main panel in Fig. 1(b) shows the power spectrum of the free running (i.e. without any locking signal) feedback oscillator obtained at magnetic field of 50 Oe along y-axis, bias current of -2.3 mA and amplifier gain of 27 dB. The same spectrum is shown in linear scale in the first inset of Fig. 1 (b). This shows that the device works as a feedback oscillator with a high quality factor (Q=frequency / line width) of 8080. One can see that the main peak is accompanied by side bands in the main panel of Fig. 1(b), which are not visible in linear scale plot. The frequency difference between the side bands is about 120 MHz, which corresponds to the round trip time delay of 8 ns. The second inset of Fig. 1(b) shows the power spectrum obtained at the same magnetic field of 50 Oe along y-axis and the same bias current (-2.3 mA) but without feedback. (The feedback line was disconnected during this measurement.) The MTJ device showed a high TMR ratio of 56% as shown in the third inset of Fig. 1(b). The TMR was measured by sweeping the magnetic field along the easy axis (X-axis). One can see from Fig. 1(b) a large increase in the quality factor and amplitude of power spectrum when feedback is connected.

Next we present detailed results on synchronization measurements. An RF signal generator was connected to the feedback path via a directional coupler as shown in Fig. 1(a). The RF current passing through CPW, creates an external rf magnetic field ($h_e$) on the free layer magnetization along x axis. The magnetic field produced by the current in CPW was calibrated by comparing the shift of the TMR loop with the passage of dc current through the CPW.

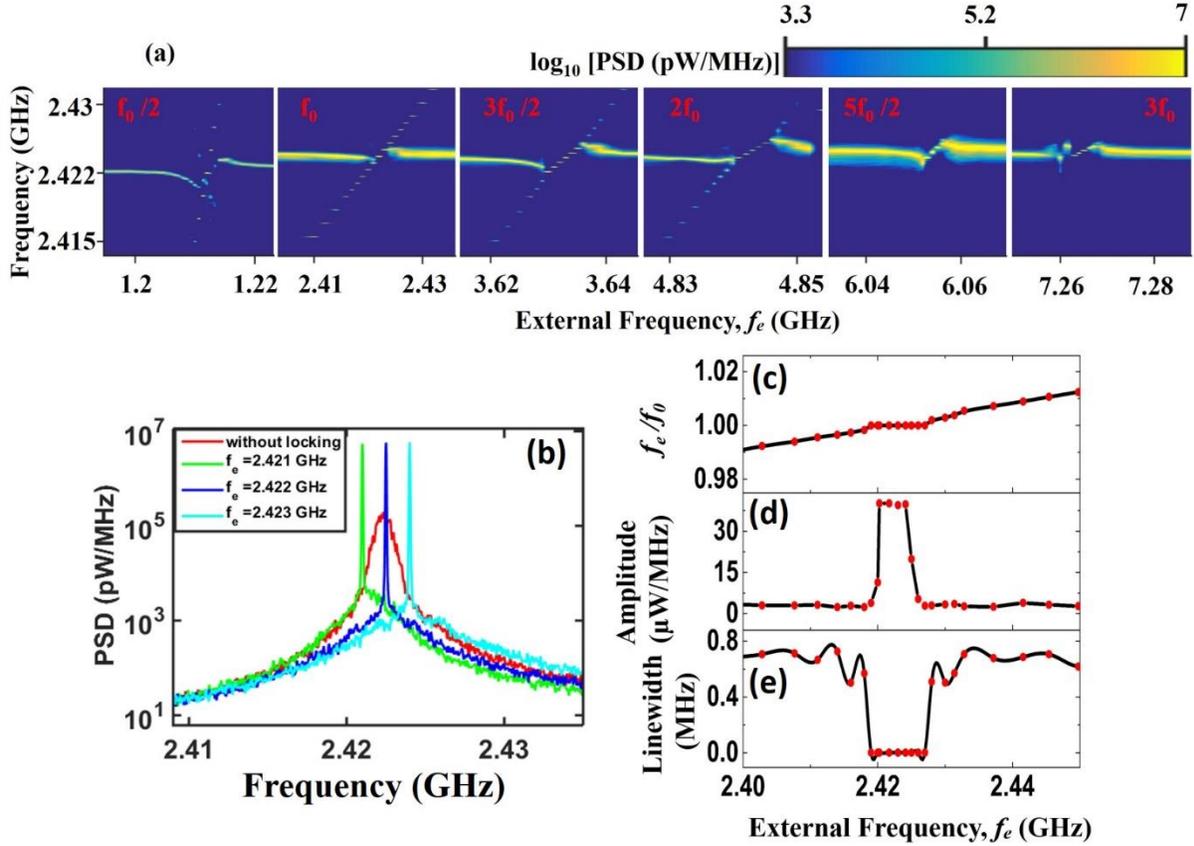

FIG. 2. (a) 2D plot of power spectral density (PSD) as a function of external frequency ($f_e$) applied at integer (n=1, 2, 3) and fractional (f=1/2, 3/2, 5/2) multiples of the auto-oscillation frequency ($f_0$) of the free running SFNO at $h_e$=7Oe. (b) PSD obtained for three values of $f_e$ close to $f_0$ (i.e. n=1 phase locking) at $h_e$=5.5 Oe. The free running PSD is also shown in the same plot for comparison. (c-e) The ratio $f_e/f_0$, peak amplitude and line width as a function of $f_e$.

Fig. 2(a) shows 2D plot of PSD as a function of external frequency ($f_e$) obtained at $h_e$=7Oe. This Figure clearly shows the phase locking phenomena observed at integer (n=1,2,3) and fractional (f=1/2, 3/2, 5/2) multiples of the free running frequency, $f_0$. Fig. 2(b) shows PSD obtained for three values of $f_e$ close to $f_0$ (i.e. n=1 phase locking). The free running PSD is also shown in the same plot for comparison. The ratio $f_e/f_0$, peak amplitude and line width as a function of $f_e$ are shown in Fig. 2(c-e). Within a certain range of external frequency, the oscillator locks to the external source and displays very small line width. The data shown in Fig. 2(a) was obtained with a bandwidth of 910 Hz. For some external frequencies, the width of the peak in the locked regime was checked by reducing the bandwidth to the minimum setting of 2 Hz, and was found to be below measurable limit.

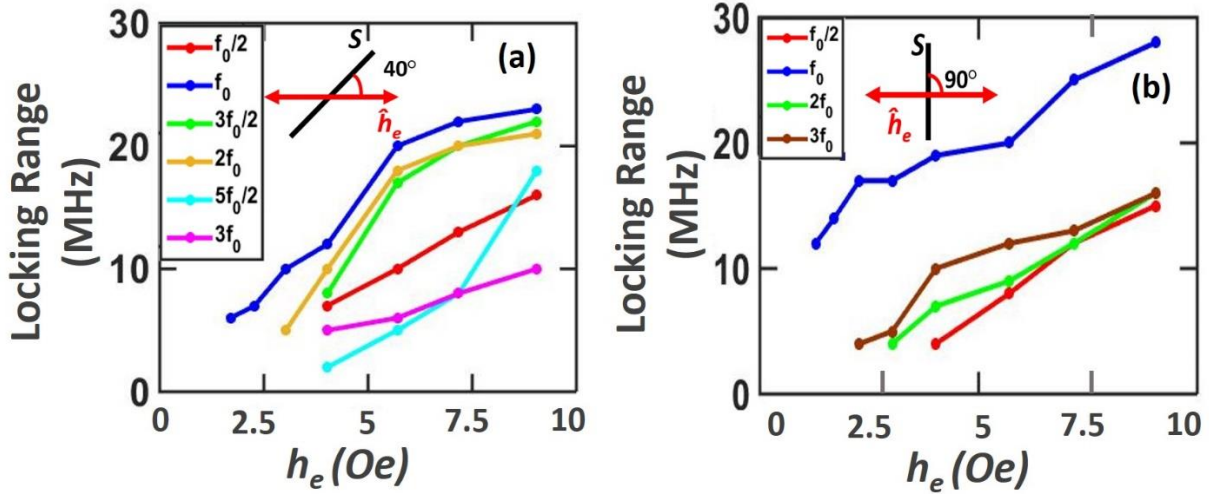

*FIG. 3. (a)* locking range *for integer and fractional multiples of free running frequency ($f_0$) as a function of $h_e$ when (a) 50 Oe field applied along Y axis (the angle between axis of oscillation (S) and* rf magnetic field ($h_e$) is $40^o$). (b) 50 Oe magnetic field applied with $135^o$ angle to Y axis (the angle between axis of oscillation (S) and rf magnetic field ($h_e$) is $90^o$).

The locking range (defined as frequency range over which oscillator frequency matches with the external frequency) is shown as a function of $h_e$ in Fig. 3(a) for integer as well as fractional multiples. We can see that the locking range is maximum for n=1 locking. It shows a linear dependence on $h_e$ for lower values, and then appears to saturate for n=1, 3/2 and 2 near $h_e$=5.5 Oe. However, the locking range is quite small compared to the case of a typical STT oscillator. According to the universal oscillator model [4], the (n=1) locking bandwidth ($\Delta$) of a non-linear auto-oscillator is given by: $\Delta \approx \sqrt{(1+\nu^2)}\, F_e/\sqrt{p_0}$, where $F_e$ denotes the amplitude of external force (proportional to magnetic field here) and $p_0$ denotes the dimensionless power of the auto oscillator. $\nu$ denotes dimensionless non-linear frequency shift given by $\nu = N/(G_+ - G_-)$, where $N$ is coefficient of non-linear frequency shift and $G_+$-$G_-$ denotes dynamic damping coefficient. The value of $N$ can be obtained from the measurement of PSD as a function of bias current and is typically few GHz in the case of STNO. In the case of SFNO, the power increases sharply with the bias current, whereas frequency depends very weakly on it. (see Fig. 2(b) and 2(c) in ref. 9). Thus the value of N is expected to be only few MHz, which results in a small value of $\nu$ and the locking bandwidth. The locking range approximated as [4], $\gamma h_e/(2\pi)$, comes out to be 27.9 MHz (for $h_e$=10 Oe, $\gamma$=2.21X10$^5$ m/As) which is comparable to the experimental results. Further, the enhanced non-linear frequency shift of STNO gives rise to a larger line width of the free running oscillator. This is because a small amplitude fluctuation results in a large phase variation if *N* is large. This is consistent with the fact that the quality factor of SFNO is much larger (~10$^4$) than typical STNO (Q<10$^3$).

The above experimental results (Fig. 3(a)) were obtained with external dc magnetic field of 50 Oe applied along y axis. The TMR loop shown in inset of Fig.1(b) shows that the $H_C$ of the free layer is 30 Oe and the fixed layer exerts a magnetic field of 35Oe on the free layer. Thus the free layer makes an angle of 40º with x axis in equilibrium when $H_y$=50 Oe is applied (See supplementary information [29] S1). This is shown schematically in the inset of Fig. 3(a): *S* denotes the axis of precession, and $\hat{h}_e$ shows the direction along which external locking magnetic field is applied. Next we applied magnetic field of 50 Oe at ~135º angle in an attempt to magnetize the free layer along the y axis. (The free layer is likely to have non-uniform equilibrium state [30], which is ignored here.) The inset of Fig. 3(b) shows that the precession axis (S) is along y direction i.e. perpendicular to $\hat{h}_e$. The locking range in this configuration is shown in Fig. 3(b). The *f*=3/2 and 5/2 locking were not observed in this configuration. The presence of fractional locking in data shown in Fig. 3(a) and its absence in Fig. 3(b) can be related to the symmetry of oscillation orbit with respect to the direction of external rf magnetic field ($h_e$) [14]. In the case of Fig. 3(b), the oscillation axis (*S*) and $\hat{h}_e$ are perpendicular, which enhances the locking band width of odd integer multiples and suppresses fractional locking. With periodic driving signal, the phase of the oscillator satisfies the equation [14]: $\dot{\phi} = 2\pi f_0 + \mu \text{Re}[g(\phi)\exp(-i2\pi f_e t)]$, where μ is the amplitude of driving force and periodic function g is determined by the oscillation trajectory and driving signal. In the later case (Fig. 3(b)), the driving force is anti-symmetric and even Fourier components of g are zero, which enhances the locking range of n=odd multiples. In the former case driving force contains both symmetric and anti-symmetric components which increases the locking bandwidth of fractional locking.

Next we explored the effect of phase locking of the main peak on the side band peaks and whether the side band peaks show phase locking phenomenon.

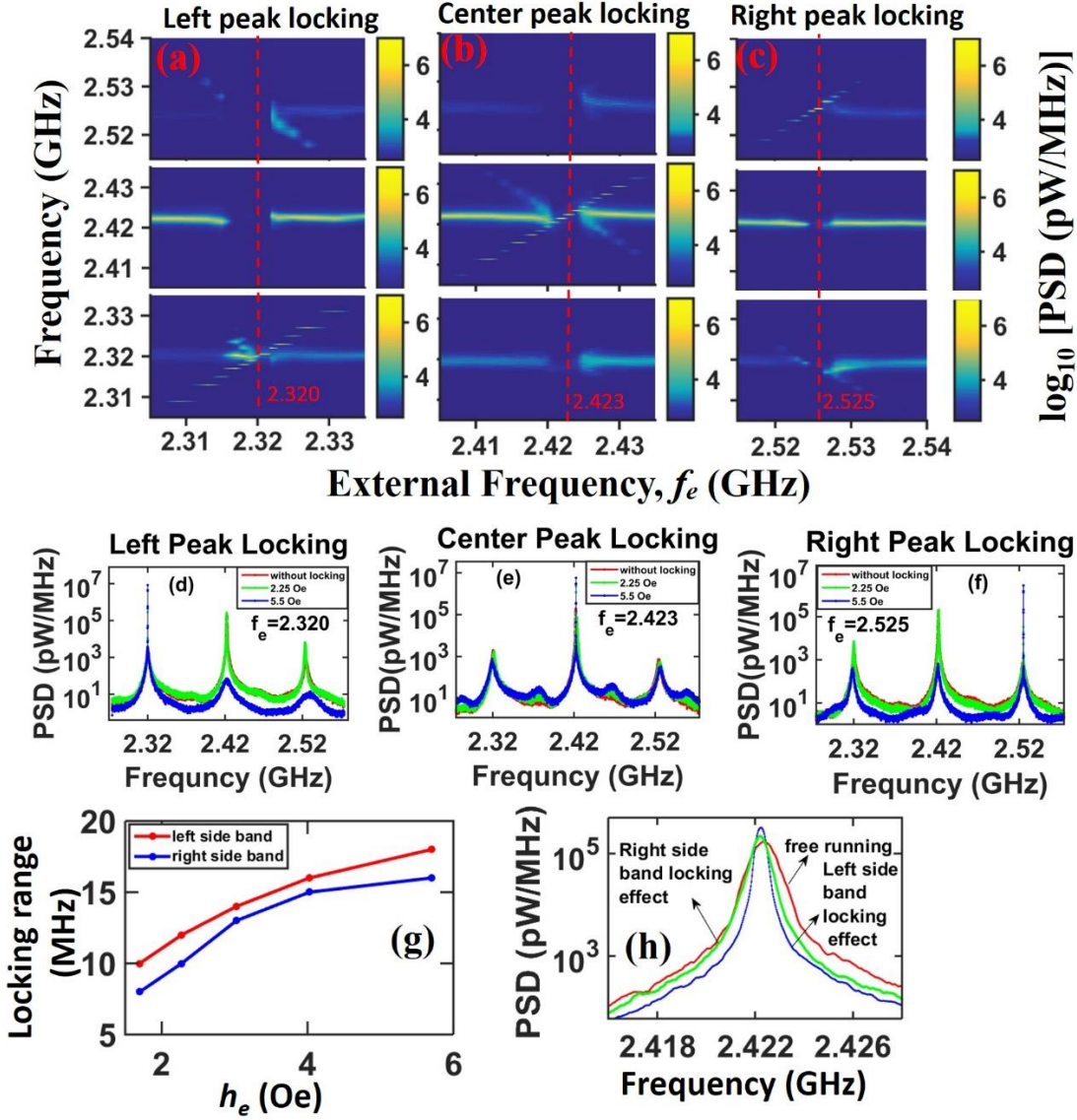

*FIG. 4. (a-c) 2D plots show locking of side band peaks and main peaks and its effect on main peak and side band peaks when locked as a function of $f_e$ at $h_e$=5.5Oe. 50 Oe magnetic field along y-axis is applied. (d-f) PSD for left side band peak, main peak and right side band peak and effect on main peak and other side band for two different values of $h_e$=2.25 Oe and 5.5 Oe. (g) Locking range of left side band and right side band as a function of $h_e$. (h) improvement in the quality factor of the main peak when side bands are locked at lower driving field ($h_e$=2.25Oe).*

The experimental results obtained at $h_e$=5.5Oe are plotted as 2D color plot data in panels a-c of Fig. 4. The panel (a) in Fig. 4 shows that when the side band peak to the left of main peak is injection locked, the main and right side band peaks are suppressed. The panel (b) in Fig. 4 shows that when the main peak is injection locked, the side band peaks are suppressed. Similar behavior was seen for injection locking of the right side band peak (panel (c) in Fig. 4). Panels (d-f) in Fig.

4 show the PSD for two values of $h_e$ (2.25 Oe and 5.5 Oe), when left side peak, main peak and the right side peak are injection locked respectively. The blue curve ($h_e$=5.5 Oe) shows the above mentioned behavior i.e. whenever one of the peaks is injection locked, the other two peaks are suppressed. Surprisingly, the behavior is quite different for low values of $h_e$. The green curve in panels d-f show PSD for $h_e$=2.25 Oe, where the above mentioned suppression of other peaks is not seen. Panel (g) in Fig. 4 show locking range of left side band and right side band as a function of $h_e$. The locking range is comparable to locking range in Fig. 3 (a) and also appears to saturate at higher rf magnetic fields. Panel (h) in Fig. 4 shows the PSD of main peak when the left and right hand side peaks are locked for low value of $h_e$=2.25Oe. The effect of left and right side band locking on PSD of main peak at higher field value are shown in supplementary information [29] Fig. S2. Here we can see that the main peak amplitude increases and line width decreases when side band peaks are locked. However, as discussed above if we increase the locking field for side band locking beyond a certain value, the main peak crashes down.

The above results show that it is possible to phase lock the SFNO even on its side band peaks. Thus two SFNOs can be mutually phase locked using side bands if the frequencies of the main peaks are quite different. The frequency difference between the main peak and side band peaks can be changed by changing the feedback delay (demonstrated in ref. 9). This provides a new way to get the side band peaks of two oscillators within locking band width. As we can see from Fig. 4 (d), when the left side band peak is injection locked, that peak now becomes the 'main' peak and other peaks look like side bands. Similarly, we can see for Fig. 4 (f), that the right side peak becomes the 'main' peak when injection locked.

The experimental results are in qualitative agreement with the macro-spin simulations of the phase locking of SFNO [15]. LLG equation was modified to include feedback magnetic field term [8].

$$\dot{\hat{m}} = -\gamma \hat{m} \times (\bar{H}_{eff} + \bar{h}_r + \bar{h}_{fb} + \bar{h}_e) + \alpha(\hat{m} \times \dot{\hat{m}})$$

Where $\hat{m}$ denotes unit vector along magnetization, $\gamma$ is the gyromagnetic ratio, $H_{eff}$ is the effective magnetic field comprising the external field and the anisotropy field, $h_r$ is the random magnetic field arising because of the thermal fluctuations, $\alpha$ is the Gilbert damping constant. $h_{fb}$ denotes the feedback magnetic field and $h_e$ denotes the external rf magnetic field. The random magnetic field satisfies the statistical properties [8]:

$$\langle h_{r,i}(t) \rangle = 0, \ \langle h_{r,i}(t) h_{r,j}(s) \rangle = 2D\delta_{ij}\delta(t-s), \ D = \alpha k_B T/(\gamma \mu_0 M_S V)$$

Where $k_B$, $T$, $\mu_0$, $M_S$ and $V$ denote the Boltzmann constant, temperature, magnetic permeability, saturation magnetization, and volume respectively. $D$ is the strength of the thermal fluctuations. The feedback field is given by [8], $h_{fb}(t) = I_{dc} \hat{x}[R(t-\Delta t) - R_0]/[2w(R_0 + R_T)]$, where $I_{dc}$ is the dc current, $w$ denotes the width of the feedback line, $R_0$ denotes the average resistance of MTJ, $R_T$ denotes the termination resistance (50 Ω). $\Delta t$ denotes the feedback delay and $R(t-\Delta t)$ is the resistance of MTJ at time $t-\Delta t$. The various parameters used in the simulation are as follows: $\alpha$=0.01, $\gamma$=2.21X10$^5$ A/ms, $T$=300 K, $M_S$=1000 emu/cc, $V$= (500 nm X 300 nm X 3nm). The anisotropy magnetic field is given by : $H_{ani} = -H_{//}m_x + H_\perp m_z$, where $H_{//}$ and $H_\perp$ denote the in-plane and out-of-plane anisotropy fields. Positive values of $H_{//}$ and $H_\perp$ imply that x-axis is the easy axis and z-axis is out-of-plane hard axis. We have used $H_{//}$=30 Oe and $H_\perp$=10$^4$ Oe. The width of the feedback strip was taken as 1µm and amplification of 20 dB was assumed.

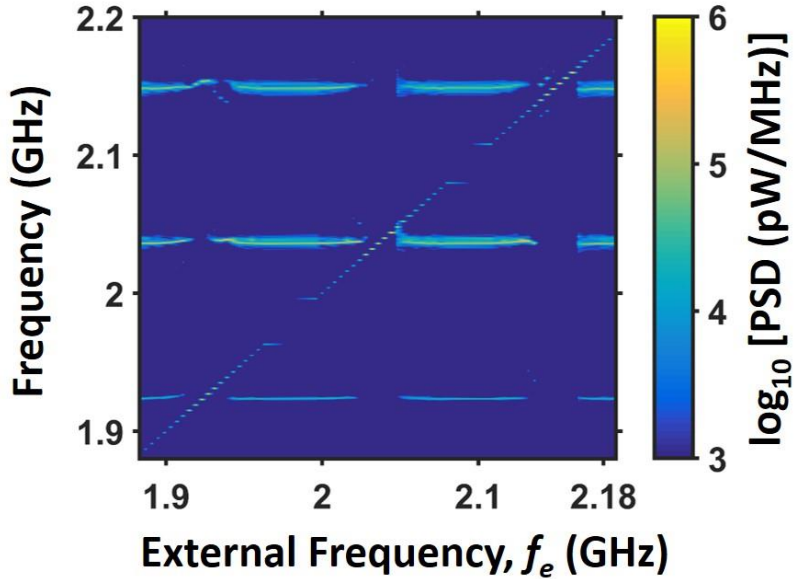

FIG. 5. Simulation results: 2D plot of PSD as a function of external frequency ($f_e$) around free running frequency ($f_o$), left side band frequency and right side band frequency with effect of locking on other peaks at $h_e$=2 Oe.

The simulation results for injection locking are shown in Fig. 5 as a 2D plot for $h_e$=2 Oe. This Figure shows that it is possible to injection lock the main peak, left and right side band peak. As observed in the experiments, simulations also show that when a peak is injection locked the other two peaks are suppressed. The simulation results for lower value of $h_e$=0.6 Oe are shown in Fig. 6 (a) as a 2D plot. Fig. 6 (b-d) show the cuts of the 2D data shown in Fig. 5 and Fig. 6 (a) for $f_e$=1.923 GHz (left side band peak locking), 2.036 GHz (center peak locking), 2.148 GHz (right side band peak locking). The free running spectrum i.e. without injection locking is also shown in Fig. 6 (b-d) for comparison.

From these Figures 6 (b-d) one can see that for higher value of $h_e$ (2 Oe), if the center or right side band peak is injection locked, the other two peaks are suppressed, If the left side band peak in injection locked, the center peak suppressed but the right side band peak is almost same as free running case. Experimentally what we see is that for higher $h_e$, if any of the three peaks is injection locked, the other two peaks are suppressed.

From Fig. 6 (b-d) one can see that for lower value of $h_e$ (0.6 Oe), if the center or left side band peak is injection locked, the other two peaks are almost the same as free running case., If the right side band peak in injection locked, the other two peaks are somewhat suppressed. Experimentally what we see is that for lower $h_e$ if any of the three peaks is injection locked, the other two peaks are not suppressed.

Thus though many aspects of the experimental results are supported by the simulations, there is some discrepancy for injection locking of left side band peak for lower $h_e$ and right side band peak for higher $h_e$. Such a mismatch could arise from the non-uniform precessional modes, which are neglected in the macro-spin simulations.

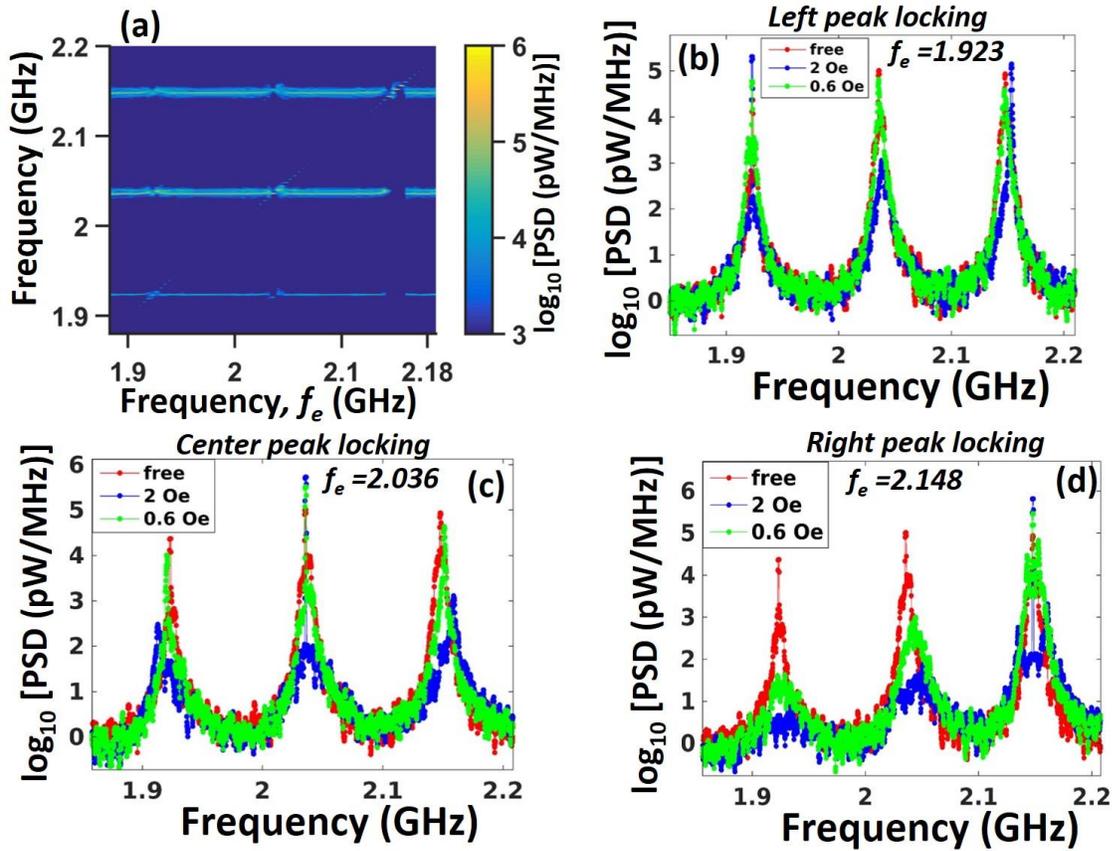

FIG. 6. Simulation results: (a) shows 2D plot of PSD as a function of external source frequency ($f_e$) around free running frequency ($f_o$), left side band frequency and right side band frequency with effect of locking on other peaks at $h_e=0.6Oe$. (b-d) PSD for left side band peak, main peak and right side band peak and effect on main peak and other side band for different values of $h_e$.

In conclusion, we have demonstrated the integer and fractional injection locking of feedback oscillator to microwave magnetic field. We explored the relation of locking range to the symmetry of oscillation orbit with respect to the direction of driving force. This was done by measuring the locking range for two different directions of the dc magnetic field. We demonstrated that feedback oscillators can be injection locked on the side band peaks. The macro-magnetic simulations are in qualitative agreement with this finding. The frequency difference between the main peak and side band peaks can be controlled by changing the feedback delay. This implies that side band locking scheme would be useful for phase locking oscillators with large frequency differences.

Acknowledgement: We are thankful to the Centre of Excellence in Nanoelectronics (CEN) at the IIT-Bombay Nanofabrication facility (IITBNF) and Ministry of Electronics and Information Technology (MeitY), Government of India for its support.

**References**

[1]. J. C. Slonczewski, Current-driven excitation of magnetic multilayers, J. Magn. Magn Mater. 159, L1 (1996).

[2]. S. I. Kiselev, J. C. Sankey, I. N. Krivorotov, N. C. Emley, R. J. Schoelkopf, R. A. Buhrman, and D. C. Ralph, Microwave oscillations of a nanomagnet driven by a spin-polarized current, Nature (London) 425, 380 (2003).

[3]. S. Sharma, B. Muralidharan, and A. Tulapurkar, Proposal for a domain wall nano-oscillator driven by non-uniform spin currents, Sci. Rep. 5, 14647 (2015).

[4]. A. Slavin and V. Tiberkevich, Nonlinear auto-oscillator theory of microwave generation by spin-polarized current, IEEE Trans. Magn. 45, 1875 (2009).

[5]. H. Kubota, K. Yakushiji, A. Fukushima, S. Tamaru, M. Konoto, T. Nozaki, S. Ishibashi, T. Saruya, S. Yuasa, T. Taniguchi, H. Arai and H. Imamura, Spin-torque oscillator based on magnetic tunnel junction with a perpendicularly magnetized free layer and in-plane magnetized polarizer. Appl. Phys. Express. 6, 103003-1-103003-3 (2013).

[6]. S. Tamaru, H. Kubota, K. Yakushiji, S. Yuasa, & A. Fukushima, Extremely Coherent Microwave Emission from Spin Torque Oscillator Stabilized by Phase Locked Loop, Sci. Rep. **5,** 18134 (2015).

[7]. A. Bose, A. K. Shukla, K. Konishi, S. Jain, N. Asam, S. Bhuktare, H. Singh, D. D. Lam, Y. Fujii, S. Miwa, Y. Suzuki, and A. A. Tulapurkar, Observation of thermally driven field-like spin torque in magnetic tunnel junctions, Appl. Phys. Lett. 109, 032406 (2016).

[8]. D. Dixit, K. Konishi, C. V. Tomy, Y. Suzuki, and A. A. Tulapurkar, Spintronic oscillator based on magnetic field feedback, Appl. Phys. Lett. 101, 122410 (2012).

[9]. D. Kumar, K. Konishi, Nikhil Kumar, S. Miwa, A. Fukushima, K. Yakushiji, S. Yuasa, H. Kubota, C. V. Tomy, A. Prabhakar, Y. Suzuki, and A. Tulapurkar, Coherent microwave generation by spintronic feedback oscillator, Sci. Rep. 6, 30747 (2016).

[10]. A. Pikovsky, M. Rosenbblum, and J. Kurths, Synchronization: A Universal Concept in Nonlinear Sciences (Cambridge, New York, 2001).

[11]. K. Roy, M. Sharad, D. Fan, and K. Yogendra, Brain-inspired computing with spin torque devices, in Design, Automation and Test in Europe Conference and Exhibition (DATE), 2014, Mar. 2014, pp. 1–6.

[12]. K. Yogendra, D. Fan, and K. Roy, Coupled Spin Torque Nano Oscillators for Low Power Neural Computation. IEEE Trans. Magn. 51, 1–9 (2015).

[13]. J. Grollier, D. Querlioz and M. D. Stiles, Spintronic Nanodevices for Bioinspired Computing. Proc. IEEE 104, 2024–2039 (2016).

[14]. S. Urazhdin, P. Tabor, V. Tyberkevych, and A. Slavin, Fractional Synchronization of Spin-Torque Nano-Oscillators, Phys. Rev. Lett. 105, 104101 (2010)

[15]. M. Quinsat, J. F. Sierra, I. Firastrau V. Tiberkevich, A. Slavin, D. Gusakova, L. D. Buda-Prejbeanu, M. Zarudniev, J.-P. Michel, U. Ebels, B. Dieny, M.-C. Cyrille, J. A. Katine, D. Mauri and A. Zeltser, Injection locking of tunnel junction oscillators to a microwave current, Appl. Phys. Lett. 98, 182503 (2011)

[16]. J. Grollier, V. Cros, A. Fert, Synchronization of spin-transfer oscillators driven by stimulated microwave currents, Phys. Rev. B,73, 060409R (2006)

[17]. V.E. Demidov, H. Ulrichs, S.V. Gurevich, S.O. Demokritov, V.S. Tiberkevich, A.N. Slavin, A. Zholud and S. Urazhdin, Synchronization of spin Hall nano-oscillators to external microwave signals, Nat. Commun. 5,3179, (2014)

[18]. A. Dussaux, B. Georges, J. Grollier, V. Cros, A.V. Khvalkovskiy, A. Fukushima, M. Konoto, H. Kubota, K. Yakushiji, S. Yuasa, K.A. Zvezdin, K. Ando and A. Fert, Large microwave generation from current-driven magnetic vortex oscillators in magnetic tunnel junctions. Nat. Commun. 1, 8-1-6 (2010).

[19]. A. Awad, P. Dürrenfeld, A. Houshang, M. Dvornik, E. Iacocca, R. K. Dumas and J. Åkerman, Long-range mutual synchronization of spin Hall nano-oscillators, Nat. Phys. 13, 292-299 (2017)

[20]. W. H. Rippard, M. R. Pufall, S. Kaka, T. J. Silva and S. E. Russek, Injection locking and phase control of spin transfer oscillators. Phys. Rev. Lett. 95, 067203 (2005).

[21]. A. Dussaux, A.V. Khvalkovskiy, J. Grollier, V. Cros, A. Fukushima, M. Konoto, H. Kubota, K. Yakushiji, S. Yuasa, K. Ando and A. Fert, Phase locking of vortex based spin transfer oscillators to a microwave current, Appl. Phys. Lett. **98**, 132506 (2011).

[22]. S. Tsunegi, E. Grimaldi, R. Lebrun, H. Kubota, A.S. Jenkins, K. Yakushiji, A. Fukushima, P. Bortolotti, J. Grollier, S. Yuasa and V. Cros, Self-Injection Locking of a Vortex Spin Torque Oscillator by Delayed Feedback, Sci. Rep. 6, 26849 (2016).

[23]. S. Kaka, M. R. Pufall, W. H. Rippard, T. J. Silva, S. E. Russek, and J. A. Katine, Mutual phase-locking of microwave spin torque nano-oscillators, Nature (London) 437, 389 (2005).

[24]. S. Bhuktare, H. Singh, A. Bose, and A. A. Tulapurkar, Spintronic Oscillator Based on Spin-Current Feedback Using the Spin Hall Effect, Phys. Rev. Applied 7, 014022 (2017)

[25]. A. Bose, S. Dutta, S. Bhuktare, H. Singh, A. Tulapurkar, Sensitive measurement of spin orbit torque driven ferromagnetic resonance detected by planar Hall geometry, Appl. Phys. Lett. 111, 162405 (2017)

[26]. M. Miron, G. Gaudin, S. Auffret, B. Rodmacq, A. Schuhl, S. Pizzini, J. Vogel, and P. Gambardella, Current-driven spin torque induced by the Rashba effect in a ferromagnetic metal layer, Nat. Mater. 9, 230 (2010).

[27]. A. Bose, H. Singh, S. Bhuktare, S. Dutta, A. Tulapurkar, Sign reversal of field like spin orbit torque in ultrathin Chromium/Nickel bi-layer, preprint arXiv:1706.07260 (2017).

[28]. Hanuman Singh et al., private communication.

[29]. See Supplemental Material at [URL will be inserted by publisher] for Calculation of equilibrium magnetization direction, Effect of left and right side band peaks locking on main peak at lower driving field and higher driving field and Calibration of magnetic field produced by the CPW.

[30]. J. Miltat and M. J. Donahue, Numerical Micromagnetics: Finite Difference Methods, Handbook of Magnetism and Advanced Magnetic Materials, Edited by H. Kronmüller and S. Parkin. Volume 2: Micromagnetism (Willey, 2007)

# Supplementary information

## Supplementary Note 1: Calculation of equilibrium magnetization direction:

A schematic diagram of the free layer magnetization along with external magnetic field is shown in the Fig. S1.

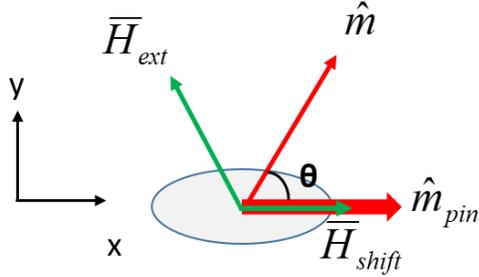

Fig S1. *Schematic diagram of the magnetization direction of the free layer and magnetic fields.*

X axis is taken as the easy axis of the free layer and y axis is taken as in-plane hard axis. An external magnetic field ($H_{ext}$) is applied in the x-y plane. The pinned layer exerts a magnetic field ($H_{shift}$) on the free layer as shown in the Fig. S1. The magnetic field acting on the free layer is given by:

$$\overline{H} = H_{//} m_x \hat{x} + \overline{H}_{pin} + \overline{H}_{ext} = (H_{//} m_x + H_{shift} + H_{ext,x})\hat{x} + H_{ext,y} \hat{y}$$

where $H_{//}$ is the in-plane anisotropy magnetic field, $H_{ext,x/y}$ are the components of the external magnetic field along x/y axes, and $m_x$ is the x component of the normalized magnetization of the free layer. The equilibrium direction of magnetization is obtained by using, $\hat{m} \times \overline{H} = 0$. This equation gives, $\cos\theta H_{ext,y} - \sin\theta(H_{//}\cos\theta + H_{shift} + H_{ext,x}) = 0$ where θ is the angle of magnetization with x-axis.

Using $H_{//}$=30 Oe and $H_{shift}$=35 Oe (determined from the TMR data), if a magnetic field of 50 Oe is applied along y-axis, we get θ≈40°. If magnetic field of 50 Oe is applied at 135° angle with x-axis, above equation gives θ≈90°.

## Supplementary Note 2: Effect of left side band and right side band peaks locking on main peak (central peak) at lower driving field ($h_e=2.25 Oe$) and higher driving field ($h_e=5.5 Oe$).

We have shown the PSD of main peak (central peak) when left side band peak and right side band peak are locked at lower driving field *($h_e=2.25Oe$)* and higher driving field *($h_e=5.5Oe$)* value in Fig. 4 of the main paper.

Fig. S2 (a-b) shows effect on PSD of main peak when side band peaks are locked at lower driving field (2.25 Oe) in linear scale and $\log_{10}$ scale respectively (Fig S2b is the same as Fig. 4h). Fig. S2 (c-d) shows effect on PSD of main peak when side band peaks are locked at higher driving field (5.5 Oe) in linear scale and $\log_{10}$ scale respectively. From these figures we can see that the quality factor and amplitude is enhanced for locking of left side band and right side band peak at lower driving field value whereas at higher driving field value main peak is suppressed.

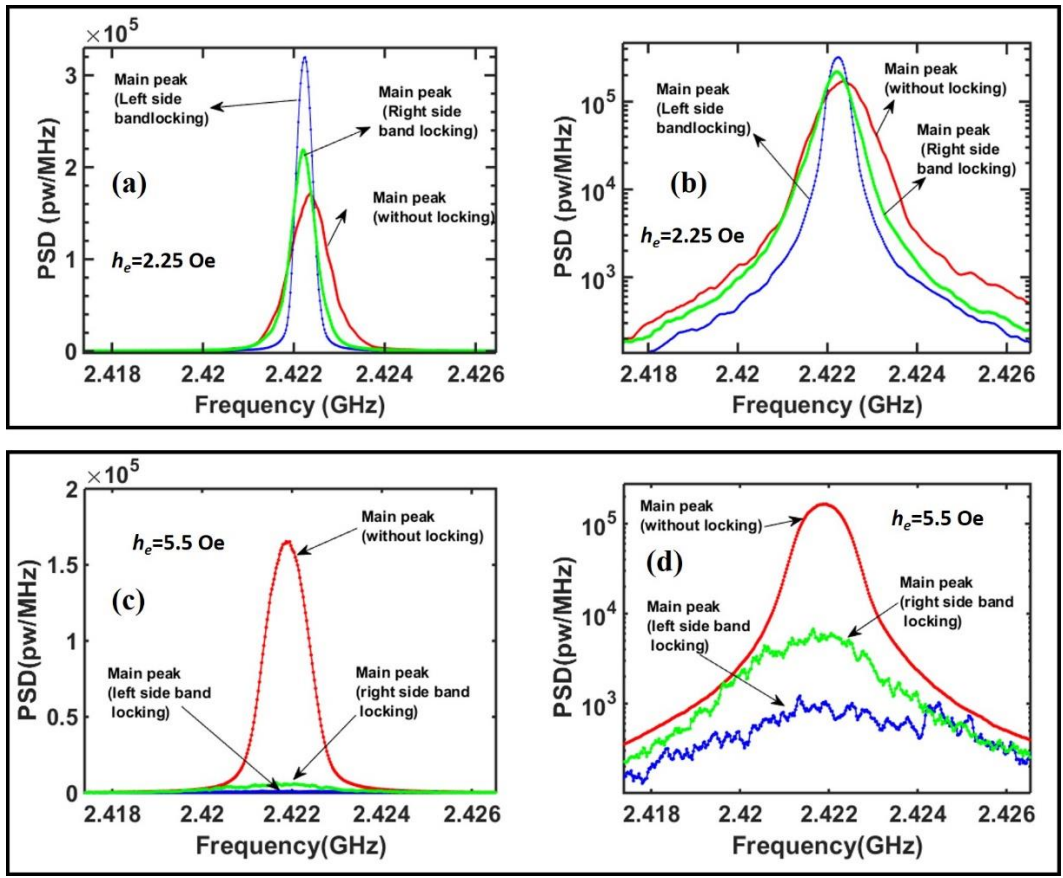

*FIG. S2. Effect on PSD of main peak (central peak) when left side band peak and right side band peak are injection locked. Fig. S2 (a) and (b) show PSD of main peak at lower driving field ($h_e=2.25Oe$) in linear and $\log_{10}$ scales respectively. Fig. S2 (c) and (d) show PSD of main peak at higher driving field ($h_e=5.5Oe$) in linear and $\log_{10}$ scales respectively. The main peak is enhance at lower driving field and suppressed at higher driving field.*

## Supplementary Note 3: Calibration of magnetic field produced by the CPW.

The TMR loops of the MTJ were measured for different values of dc currents passing through the CPW. The results are shown in Fig. S3. The x-axis of Fig. S3 shows the magnetic field produced by the electromagnet. The magnetic field produced by the dc current passing through the CPW gives rise to the shift of the loop. (Shift in the opposite direction was observed for negative values of dc currents.) From the Fig. S3 we can estimate that a current of 1 mA gives rise to a magnetic field of ~5 Oe. The MTJ used for these measurements is nominally identical to the MTJ on which injection locking experiments were done.

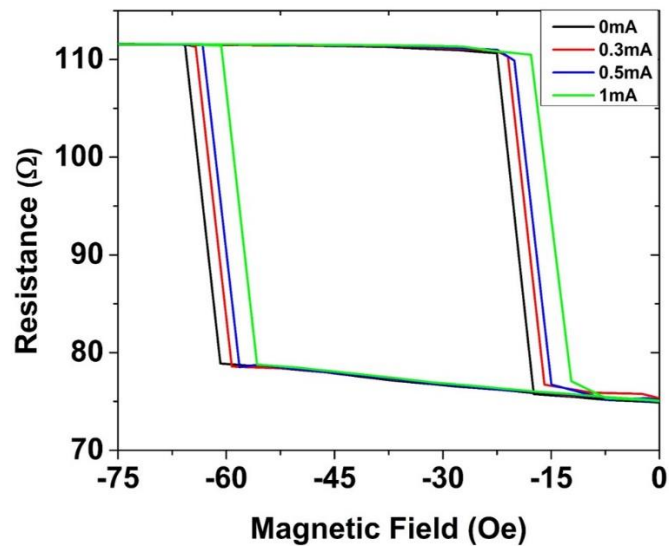

*FIG. S3.: shifting of TMR loop for different dc current through CPW.*